\documentclass[10pt,letterpaper]{article}

\usepackage{setspace} 
\usepackage{amsmath} 
\usepackage[version=4]{mhchem}	

\usepackage{rotating}
\usepackage{graphicx}
\usepackage[export]{adjustbox}
\usepackage[aboveskip=1pt,labelfont=bf,labelsep=period,singlelinecheck=off]{caption}
\usepackage{graphicx}
\usepackage{sidecap}
\usepackage{wrapfig}
\usepackage[pscoord]{eso-pic}

\usepackage[left]{lineno}
\usepackage{microtype}
\DisableLigatures[f]{encoding = *, family = * }
\usepackage[numbers,sort&compress,square]{natbib}
\usepackage[utf8]{inputenc}
\usepackage{nameref,hyperref}
\makeatletter
\renewcommand{\@biblabel}[1]{\quad#1.}
\makeatother

\usepackage{changepage}
\usepackage[datesep={},timesep={}]{datetime2}
\usepackage{pdfpages}
\usepackage{lastpage,fancyhdr}
\usepackage{epstopdf}
\usepackage{color}
\definecolor{red}{cmyk}{0,1,1,0}

\usepackage[fulladjust]{marginnote}
\usepackage{booktabs}

\reversemarginpar

\begin{document}
\vspace*{0.35in}

\begin{flushleft}
{\Large
\textbf\newline{Directional aspects of vegetation linear and circular polarization biosignatures}
}
\newline
\\
C.H. Lucas Patty\textsuperscript{1*},
Antoine Pommerol\textsuperscript{1},
Jonas G. K\"uhn\textsuperscript{2,3},
Brice-Olivier Demory\textsuperscript{2},
Nicolas Thomas\textsuperscript{1}

\bigskip
\tiny 1 Physikalisches Institut, Universit\"at Bern, CH-3012 Bern, Switzerland
\\
{2} Center for Space and Habitability, Universit\"at Bern,  CH-3012 Bern, Switzerland
\\
{3} D\'epartement d'Astronomie, Universit\'e de Gen\`eve, CH-1290 Versoix, Switzerland
\\

\bigskip

\textbf{*lucas.patty@unibe.ch, chlucaspatty@gmail.com}

\end{flushleft}

\section*{Abstract}
Homochirality is a generic and unique property of all biochemical life and is considered a universal and agnostic biosignature. Upon interaction with unpolarized light, homochirality induces fractional circular polarization in the light that is scattered from it, which can be sensed remotely. As such, it can be a prime candidate biosignature in the context of future life-detection missions and observatories. The linear polarizance of vegetation is also sometimes envisaged as a biosignature, although it does not share the molecular origin as circular polarization. It is known that the linear polarization of surfaces is strongly dependent on the phase angle. The relation between the phase angle and circular polarization stemming from macromolecular assemblies, such as in vegetation, however, remained unclear. We demonstrate in this study using the average of 27 different species that the circular polarization phase angle dependency of vegetation induces relatively small changes in spectral shape and mostly affects the signal magnitude. With these results we underline the use of circular spectropolarimetry as a promising agnostic biosignature complementary to the use of linear spectropolarimetry and scalar reflectance. 
\\
\textbf{Spectropolarimetry, Biosignatures, Remote Sensing, Homochirality, Photosynthesis, Life Detection}
\section{Introduction}

While in recent decades the use of polarimetry for aerosol particle and cloud characterization has rapidly been established, optical remote sensing of surfaces using polarimetry is not broadly established. Instrumentation for the acquisition of sensitive linear polarimetry has matured towards the point that they can be readily applied for such purposes. Interpretation of the linear polarizance, however, is difficult especially due to the strong phase angle dependency. 

In recent years, circular spectropolarimetric instruments have also been developed allowing the sensitive acquisition of circular polarization of vegetation in the field \citep{Patty2019} and from a swift moving helicopter \citep{Patty2021}. The circular polarizance by biopigments provides a unique spectropolarimetric biosignature and is generally absent in the light scattered of abiotic surfaces \citep{Patty2019, Patty2021, Sparks2009}. While abiotic materials can also create circular polarization, such as through multiple scattering by clouds or aerosols, these interactions result in a weaker and spectrally flat circular polarization signal \citep{Rossi2018}. Rather than being the result of surface scattering, the circular polarizance of photosynthetic organisms, including algae and bacteria, primarily results from interactions with chiral pigment-protein complexes inside the cells \citep{Patty2018a, Sparks2021, Patty2018c}. 

Terrestrial biochemistry is based on chiral molecules which determine both the functioning and structure of biological systems. Unlike abiotic chemistry, where these molecules occur in equal, racemic, concentrations of both versions, living nature almost \citep{Grishin2020} exclusively utilizes these molecules in only one of the mirror image configurations (homochirality). Biological macromolecules and biomolecular architectures are often also homochiral. The $\alpha$-helix, for example, the most prevalent secondary structure of proteins, is almost \citep{Novotny2005} exclusively right-hand-coiled. Homochirality is a prerequisite for self-replication and thus for life as we know it \citep{Popa2004, Bonner1995, Jafarpour2015}. It is likely that homochirality is a universal feature of life  \citep{Wald1957,MacDermott2012,Patty2018a,Sparks2009} and may serve as an agnostic trait of life itself. Nonetheless, a lack of homochirality does not necessarily indicate an absence of life and false-negative scenarios can be possible \citep{Avnir2021}.

The chirality of molecules and molecular architectures endows them a specific response to electromagnetic radiation \citep{Fasman2013,Patty2018a}. When unpolarized incident light, such as emitted from the Sun \citep{Kemp1987}, is reflected off a homochiral material it induces a wavelength dependent fractional circular polarization signal, which can thus be detected remotely. It has been demonstrated that the circular polarizance of vegetation carries the same information as circular dichroism \citep{Patty2017,Patty2018b}, which is the differential absorbance of left- and right-handed circularly polarized incident light, and has has proven to be an indispensable tool in biomolecular research. 

In vegetation, the chlorophylls are organized in chiral protein-pigment aggregates. It is the chirality of the supramolecular structure, that dominates the circular polarization spectrum rather than the chirality of the constituent molecules themselves. While the signals resulting from smaller scale chirality are superimposed, the total polarization spectrum is not the sum of these spectra alone \citep{Keller1986, Barzda1994}. These supramolecular chiral aggregates can create very intense polarization signals (polymer and salt induced (psi-type) polarization) that can be up to two orders of magnitude larger than the polarization of the constituents. Additionally, psi-type polarization is characterized by non-conservative anomalously shaped bands that range beyond the wavelengths of molecular absorbance. Due to the direct relation between the circular polarizance and the macromolecular structure it is an highly sensitive indicator of the membrane macro-organization of the photosynthetic apparatus and its functioning \citep{Garab1996,Garab2009}. 

As such, vegetation circular polarizance could provide sensitive monitoring of vegetation physiology \cite{Lambrev2019}. It has been demonstrated that circular polarization can be very indicative of stress factors such as drought and can in principle also be used to detect different physiological stress responses \citep{Patty2017, Lambrev2019}. While vegetation circular polarizance is typically less than 1\%, making its use challenging for instrumentation, polarization remote sensing could prove to be a very powerful complementary tool in assessing climate change, desertification, deforestation and the general monitoring of vegetation physiological state in vulnerable regions \citep{Patty2021}.

Additionally, also linear dichroism spectroscopy has contributed significantly to the understanding of molecular organization. In principle, linear polarizance would yield the same results as linear dichroism. However, the retrieval of structural information through linear spectropolarimetry is dependent on the molecular alignment of the sample. In a randomly oriented sample this information is therefore averaged out, even if the sample possesses intrinsically anisotropic molecular architectures \citep{Garab1996,Garab2009}. The linearly polarized proportion of the light reflected of vegetation generally stems from leaf surface polarization and the light scattered by leaf internal structures is only polarized to a small extent \cite{Vanderbilt1985a, Grant1987}. As such, linear polarizance will not carry any information on molecular arrangements. It has however been demonstrated that remote sensing using linear polarization can be informative of plant physiological status \citep{Vanderbilt2017,Vanderbilt2019,Grant1993}, and can aid in estimating leaf chlorophyll content \citep{Yao2020}. The linear polarizance of vegetation can be a sensitive biosignature \cite{Berdyugina2016, Klindzic2021} that can be complementary to the scalar reflectance and circular polarizance.

Beside leaf optical properties such as the optical index of refraction from the cuticle and/or epidermis and the surface roughness, the linear polarizance of leaves is highly dependent on the angle of the incident light \citep{Vanderbilt1991,Vanderbilt1985a}. This is largely due to the increase in specular reflection by the leaf surface as a function of the angle of incidence \citep{Peltoniemi2015}. This effect becomes stronger in the pigment absorption bands, where the main contribution to the total reflection comes from the specular reflectance on the surface and the scattered light by the leaf internal structures will be absorbed. Reciprocally, the linear polarizance is lowest in the near-infrared, where no significant absorption occurs and the contribution of internally scattered unpolarized light is the greatest.  

The relation between the circular polarizance of vegetation and the angle of incidence and leaf optical properties has not been elucidated. In the past, various studies have investigated the circular polarization scattering on a molecular scale, indeed the circular polarizance is not only influenced by differential absorption but also by differential scattering. Circular Intensity Differential Scattering (CIDS) confirmed the existence of chirally organized macrostructures which show extraordinary large polarizance if the helical pitch and diameter commensurate with the wavelength \cite{Garab1988a, Garab1988b,Bustamante1985,Bustamante1983}. Also in leaves, this effect was thought to play an important role on small scale circular polarizance, i.e. leading to different polarization signals for normal leaf tissue and veins \citep{Patty2018b}. As of yet, however, no systematic experimental study on the effect of incident angles on vegetation circular polarizance has been carried out. While some studies demonstrate the possibilities to detect the circular polarizance in the laboratory or in the field, these studies have either used an integrating sphere or did not record the angles of incidence and reflection (e.g. \citep{Sparks2009, Patty2019, Patty2021}).

In the current study we present, to the best of our knowledge, for the first time the systematic evaluation of the relation between the angle of incidence and the full-Stokes polarizance of leaves. A better understanding of this phenomenon will be crucial in the further development of possible future remote sensing applications using circular polarizance and will ultimately aid in interpreting spectropolarimetric signals in future life-detection missions. 

\section{Materials and methods}
\begin{figure}[!hb]
	\centering 
	\hspace*{0cm}  
	\includegraphics[width=0.58\textwidth]{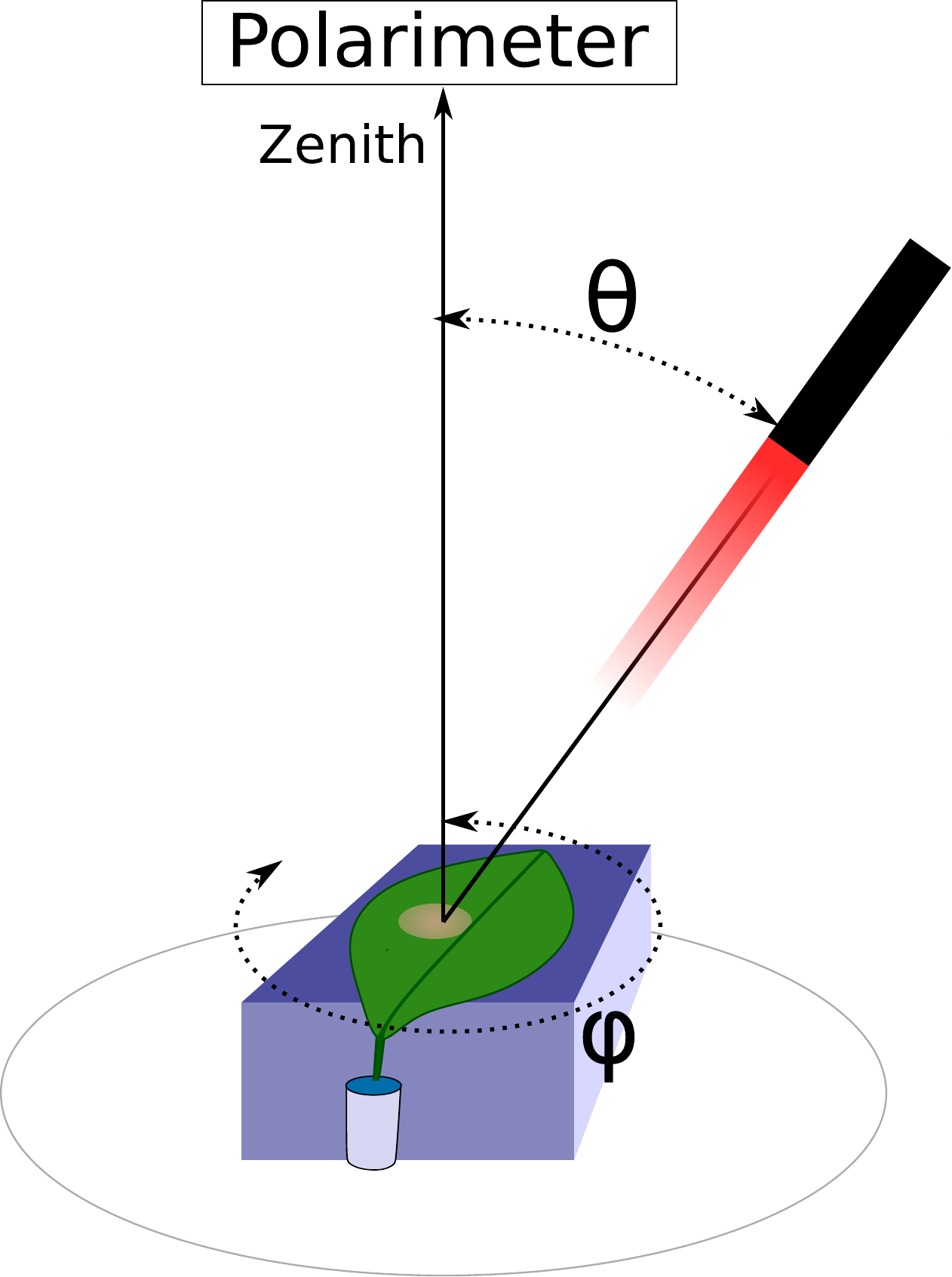}
	\caption{Schematic of the measurement setup. The detector always detects at nadir direction. $\theta$ represents the phase angle, $\varphi$ represents the azimuthal angle.}
	\label{fig:Setup}
\end{figure}
The polarimeter (Hinds Instruments II/FS42-47, USA) is based on dual photoelastic modulator (PEM) modulation and is able to record the full normalized Stokes vector with minimal risk of systematic crosstalk. It consists of 2 PEMs oriented at 45$^\circ$ modulated at 42 and 47 kHz resonance frequencies, a prism analyzer that is 22.5$^\circ$ to the PEMs and a photomultiplier tube (Hamamatsu R955, Japan). 

Polarization in general is described in terms of the Stokes vector $\textbf{S} = (I,Q,U,V)$, where $I$ is the absolute intensity, $Q$ and $U$ are the unnormalized linear polarization parameters and $V$ is the unnormalized circular polarization parameter. We describe the complete polarization states in terms of the normalized Stokes parameters $Q/I$, $U/I$ and $V/I$, where $Q/I$ denotes the difference between linear polarization normal and parallel to the plane of scattering, $U/I$ denotes the difference but with a 45$^\circ$ offset and $V/I$ denotes the difference between right-handed and left-handed circularly polarized light. A value of 1 or -1 indicates 100\% polarized light.

Light from a 100 watt halogen light source was fed to a monochromator (Oriel instruments MS257, USA), with a FWHM of 10 nm, which is consequently fiber launched and collimated and depolarized in a head mounted on a rotating arm. The polarimeter was always detecting at nadir direction. The angle of the incident light, which is similar to the phase angle $\theta$, varied between 10-75$^\circ$ with an increment of 2.5-7.5$^\circ$ varying per sample. The measured area was approximately 2 cm in diameter.

The leaves measured in this study were all picked from vegetation growing in and around the city of Bern, Switzerland, or are taken from a private indoor collection (see also Table \ref{tab:plants} for a list of the species). The vegetation species used in this study do not represent any biome. The leaves were rinsed and padded dry prior to each measurement. The leaves where flattened and attached to a rotation stage which can change the azimuth angle $\varphi$ by  $0-360^\circ$ and was generally sampled in steps of 45$^\circ$. All leaves were measured with the apex at $\varphi =0^\circ$ and where applicable the leaf surface left of the midrib was measured. The petioles were placed in a small tube containing water in order to prevent dessication during the measurements, which had a duration of 24-72 hours per leaf depending on the measurement parameters.

\begin{table}[ht]
	\centering
	\caption{The measured vegetation species.}
	\begin{tabular}[t]{lcc}
		\toprule
		&Vegetation species&\\
		\midrule
\textit{Acer platanoides}  & \textit{Ficus maclellandii}  & \textit{Prunus lusitanica}   \\
\textit{Allium ursinum}    & \textit{Hedera helix}        & \textit{Quercus robur}       \\
\textit{Buddleja davidii}  & \textit{Hosta} sp. 		  & \textit{Reynoutria japonica} \\
\textit{Calathea makoyana} & \textit{Hosta crispula}      & \textit{Rosa} sp.            \\
\textit{Capsicum chinense} & \textit{Hosta plantaginea}   & \textit{Rumex obtusifolius}  \\
\textit{Castanea sativa}   & \textit{Lonicera maackii}    & \textit{Schefflera actinophylla}\\
\textit{Cornus alba}       & \textit{Mandevilla sanderi}   & \textit{Scindapsus pictus}   \\
\textit{Dypsis lutescens}  & \textit{Prunus avium}        & \textit{Syringa vulgaris}    \\
\textit{Epimedium alpinum} & \textit{Prunus laurocerasus} & \textit{Viola odorata}      \\
		\bottomrule
	\end{tabular}
	\label{tab:plants}
\end{table}%

\begin{figure}[!htb]
	\centering 
	\hspace*{0cm}  
	\includegraphics[width=1.05\textwidth]{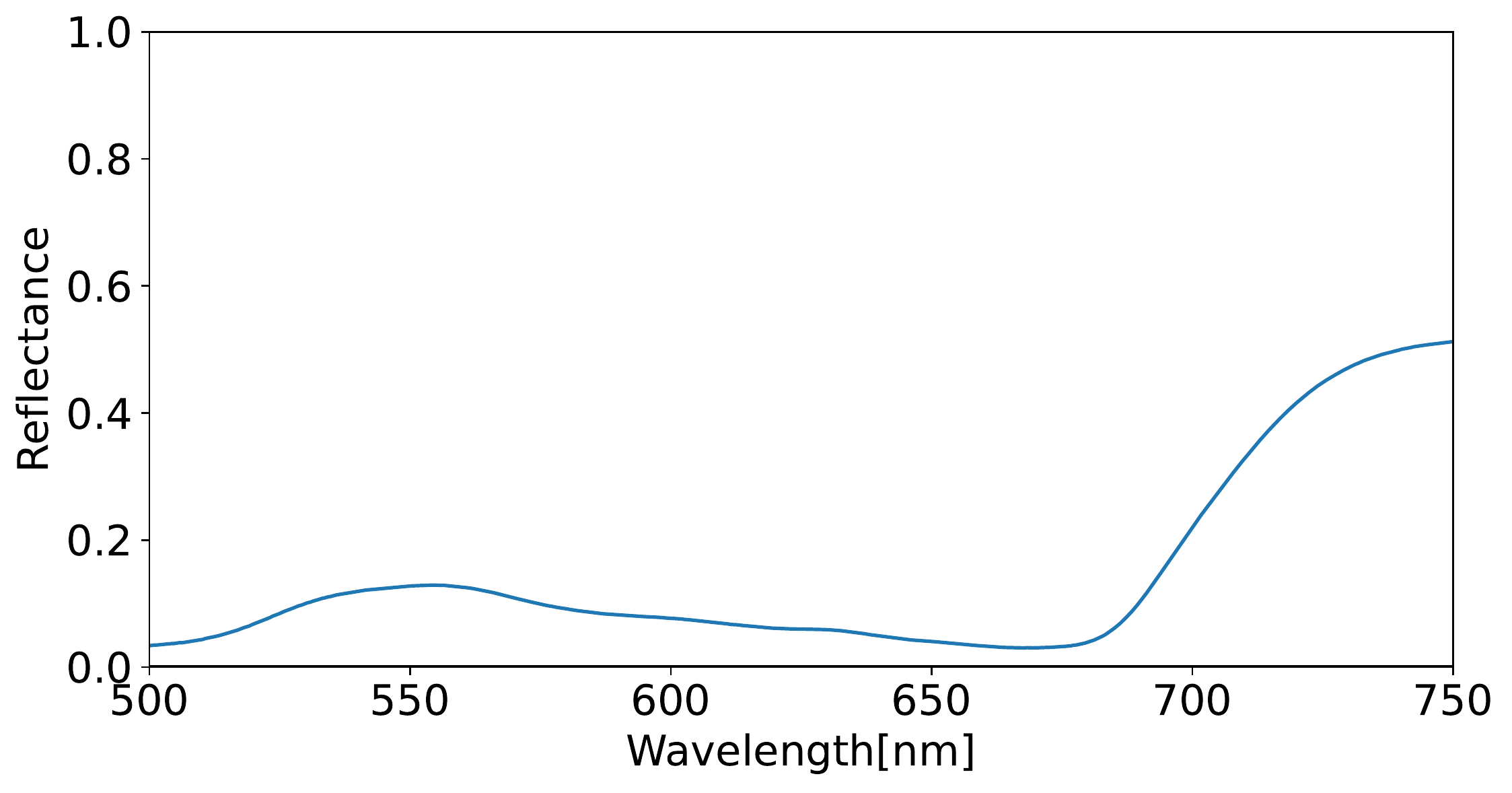}
	\caption{Typical vegetation reflectance spectrum for comparison.}
	\label{fig:Reflect}
\end{figure}

As Stokes $I$ was not recorded quantitatively in this study, we present a typical vegetation reflectance spectrum in Figure \ref{fig:Reflect}, to allow for comparison between the location of the pigment absorption bands and the presented polarizance data.

\section{Results}
\subsection{Species}

\begin{figure}[!htb]
	\centering 
	\hspace*{0cm}  
	\includegraphics[width=1.05\textwidth]{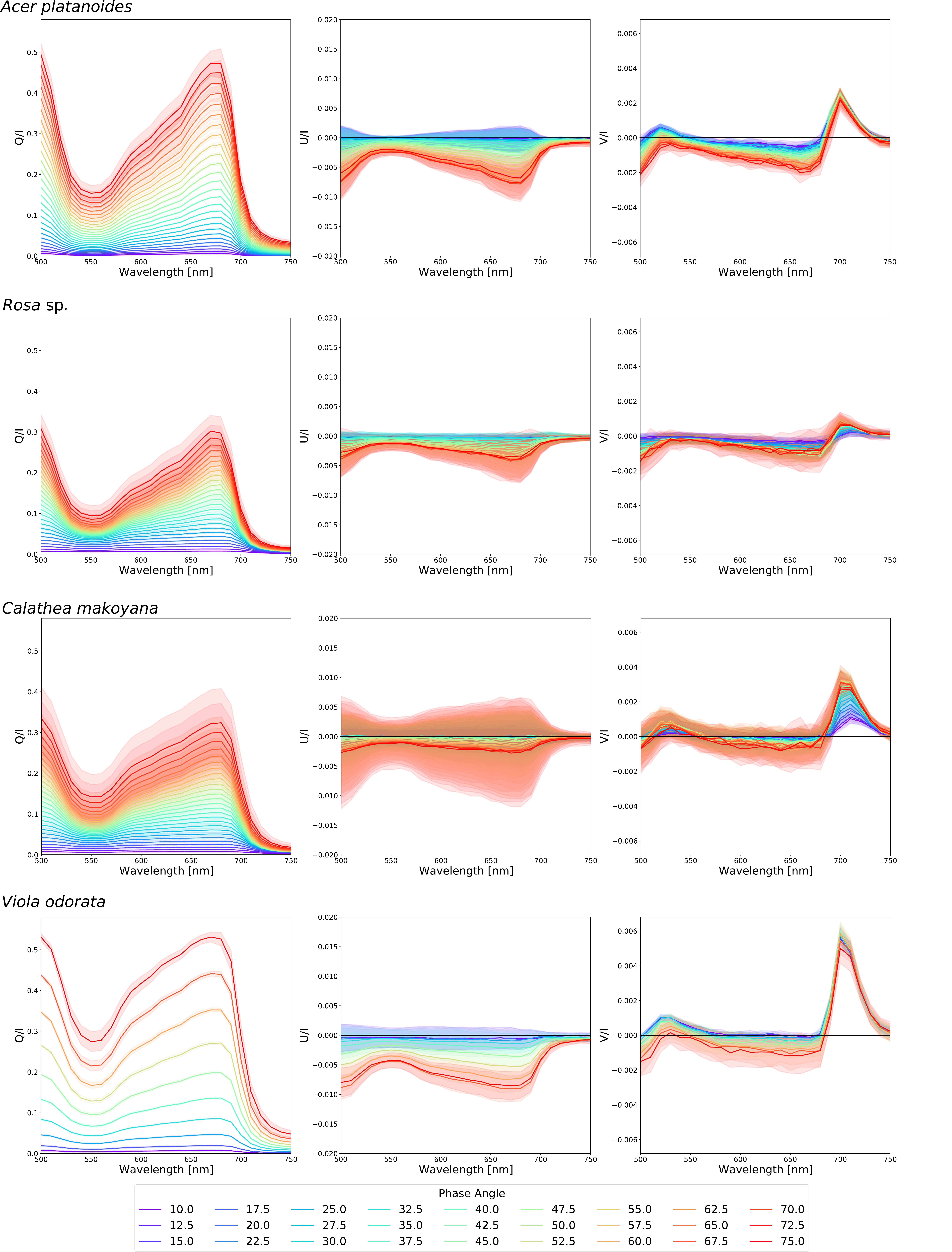}
	\caption{The polarization spectra $Q/I$, $U/I$ and $V/I$ of \textit{Acer platanoides}, \textit{Rosa} sp., \textit{Calathea makoyana} and \textit{Viola odorata} averaged over azimuth. The colors represent the different phase angles and the shaded areas denote the standard deviation.}
	\label{fig:Species}
\end{figure}

Multiple different species were measured (see also Table \ref{tab:plants}) in this study. We show some representative full-Stokes polarization spectra of some of these species in Figure \ref{fig:Species} and present the average of 27 species as the vegetation polarization biosignature in the rest of this study. The averaged full-Stokes polarization spectra of all the leaves is shown in Figure \ref{fig:Stokesgraphim}, while the polarization spectra of all individual leaves is shown in the appendix. Differences between the species can be observed, most notable in terms of polarization magnitude. For instance, \textit{Rosa} sp. has both a very low $Q/I$ and $V/I$ even at high phase angles, with a maximum $Q/I$ of $0.30$ at a phase angle of $75^\circ$ and $V/I$ of $9.1*10^{-4}$ at a phase angle of $60^\circ$. On the other hand, \textit{Viola odorata} has a much higher linear and circular polarizance, with a $Q/I$ of $0.53$ at a phase angle of $75^\circ$ and a $V/I$ of $5.9*10^{-3}$ at a phase angle of $45^\circ$. 

For some species such as \textit{Acer platanoides} and \textit{Viola odorata} the positive psi-type circular polarization band remained virtually unchanged with changes in measurement geometry, whereas this effect was much larger for instance for \textit{Rosa} sp. and \textit{Calathea makoyana}. Contrarily, the phase angle dependency of the negative psi-type band was very small for \textit{Calathea makoyana} but much larger for \textit{Acer platanoides}. Important here is the difference between these two bands, which we describe as the polarization edge below.

\subsection{Effects of the phase angle}

\begin{figure}[!htb]
	\centering 
	\hspace*{-3cm}  
	\includegraphics[width=1.5\textwidth]{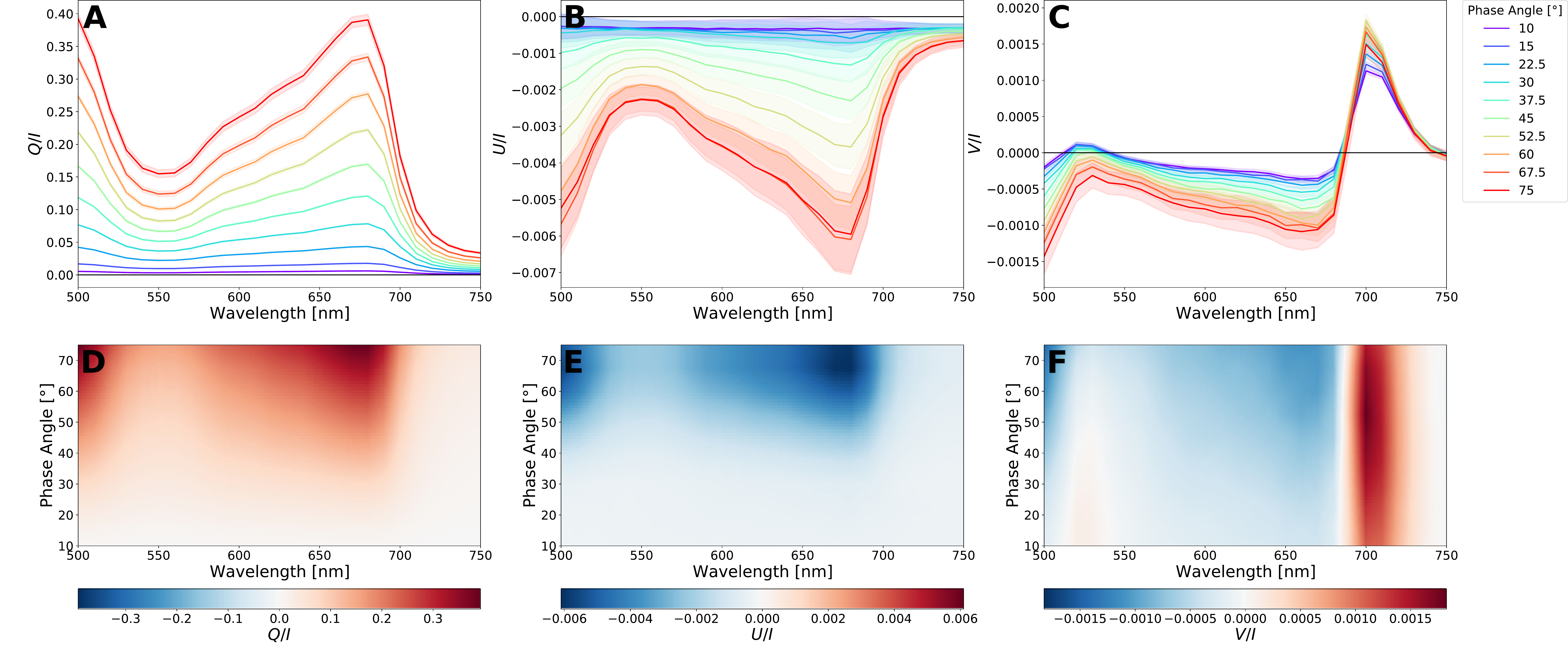}
	\caption{The average polarization spectra $Q/I$ (\textbf{A},\& \textbf{D}) $U/I$ (\textbf{B} \& \textbf{E}) and $V/I$ (\textbf{C} \& \textbf{F}) of 27 leaves of different species averaged over azimuth. In the upper panels, the colors represent the different phase angles and the shaded areas denote the standard error.}
	\label{fig:Stokesgraphim}
\end{figure}

\begin{figure}[!ht]
	\centering 
	\hspace*{-3cm}  
	\includegraphics[width=1.5\textwidth]{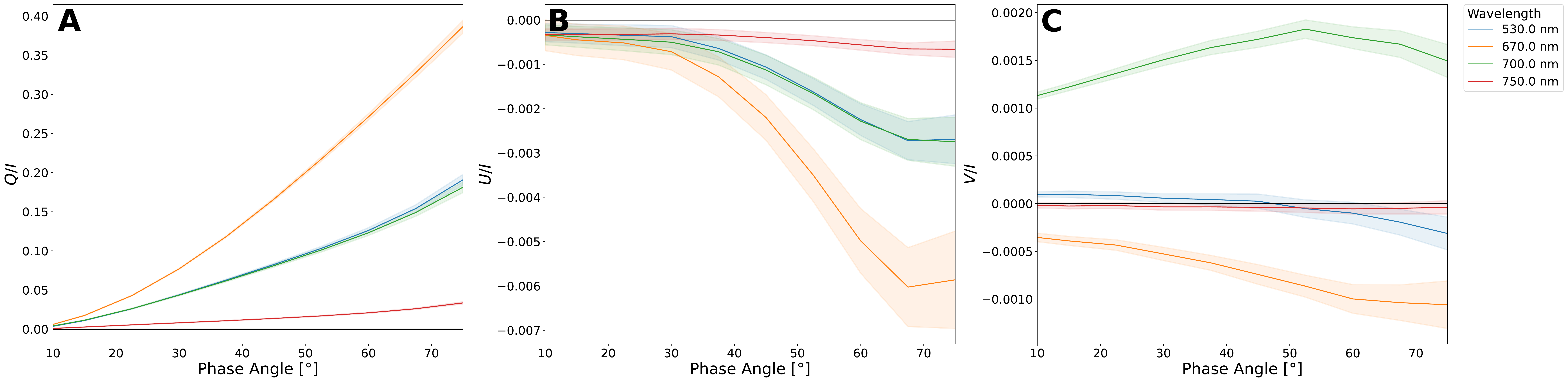}
	\caption{The polarizance, $Q/I$(\textbf{A}), $U/I$(\textbf{B}) and $V/I$ (\textbf{C}), versus phase angle for selected wavelengths. Shaded areas denote the standard error.}
	\label{fig:Stokesmaxmin}
\end{figure}

We show in Figure \ref{fig:Stokesgraphim} (\textbf{A},\textbf{B} and \textbf{C}) the average linear and circular polarization spectra ($Q/I$, $U/I$ and $V/I$ respectively) of 27 different leaves per phase angle. Figure \ref{fig:Stokesgraphim} (\textbf{C},\textbf{D} and \textbf{E}) represents the same data visualized in a 2-D display. Selected wavelengths versus the phase angle are shown in Figure \ref{fig:Stokesmaxmin} for $Q/I$ (\textbf{A}), $U/I$ (\textbf{B}) and $V/I$ (\textbf{C}). The wavelengths that were selected are the maximum and minimum magnitude of the linear polarizance (670 nm and 750 nm)and the maximum magnitudes of the positive psi-type band in the blue (530 nm) and the positive and negative psi-type bands in the red (670 nm and 700 nm respectively). 

For all wavelengths, \textit{Q/I} gradually increases with increasing increment angles (see also Figure \ref{fig:Stokesgraphim} (\textbf{A} and \textbf{D}) and Figure \ref{fig:Stokesmaxmin} (\textbf{A})). The spectral features of $Q/I$ closely resembles the inverse scalar reflectance spectrum, with low polarizance in the green (550 nm) and in the near-infrared ($>700$ nm). Additionally, the linear polarizance was especially large around the chlorophyll absorption band. This general inverse relation between the linear polarization and the albedo is consistent with the Umov effect \cite{Umov1905} and results from the attenuation of multiple scattered light in the leafs interior (see also Figure \ref{fig:Reflect} for the general vegetation reflectance). For 670 nm, the wavelength where the linear polarizance has the largest amplitude, the polarizance increased from on average $6*10^{-3}$ (or 0.6\%) at a 10$^\circ$ inclination angle to a polarizance of 0.39 (or 39\%) at a phase angle of 75$^\circ$. For 750 nm, this increased from $9*10^{-4}$ (0.09\%) to 0.03 (3\%). 

The normalized Stokes parameter \textit{U/I} shows a different spectral dependency on the phase angle (Figure \ref{fig:Stokesgraphim} (\textbf{B} and \textbf{E}) and Figure \ref{fig:Stokesmaxmin} (\textbf{B}))). Essentially, the $U/I$ appears similar to $Q/I$ at larger phase angles. At 670 nm, where $U/I$ has the largest magnitude, the polarizance is $3*10^{-4}$. Unlike $Q/I$, however, $U/I$ stays relatively small ($7*10^{-4}$) and free of any spectral features until a phase angle of $30^\circ$. Between $30^\circ$ and $67.5^\circ$ the polarizance gradually increases, reaching a maximum value of $5.9*10^{-3}$ after which it flattens out. 

The circular polarizance \textit{V/I} is in comparison with the linear polarizance ($Q/I$ and $U/I$) much less sensitive to changes in the phase angle, as can be seen in Figure \ref{fig:Stokesgraphim} (\textbf{C} and \textbf{F}) and Figure \ref{fig:Stokesmaxmin} (\textbf{C}). A small positive band can be observed around $\sim $510-540 nm (Figure \ref{fig:Stokesgraphim} (\textbf{C} and \textbf{F})), which is the psi-type band in blue stemming from the macro-organization of the pigment system. Two more psi-type bands can be observed: a negative band with the largest magnitude around 670 nm and a positive band with a peak magnitude around 700 nm. Interestingly, while the positive band has its highest magnitude around $52.5^\circ$, with an amplitude of $1.9*10^{-3}$ a similar response is not observed for the negative band. The amplitude of the negative band flattens out after $60^\circ$ with a maximum amplitude of $-1.1*10^{-3}$. Typically the blue psi-type circular polarizance band is much smaller than the psi-type bands at longer wavelengths, as can also be seen in our data. While the band keeps a similar spectral shape, the whole band increasingly becomes more negative with increasing phase angles.

\subsection{Effects of the azimuthal rotations}

\begin{figure}[!b]
	\centering 
	\hspace*{-1.5cm}  
	\includegraphics[width=1.2\textwidth]{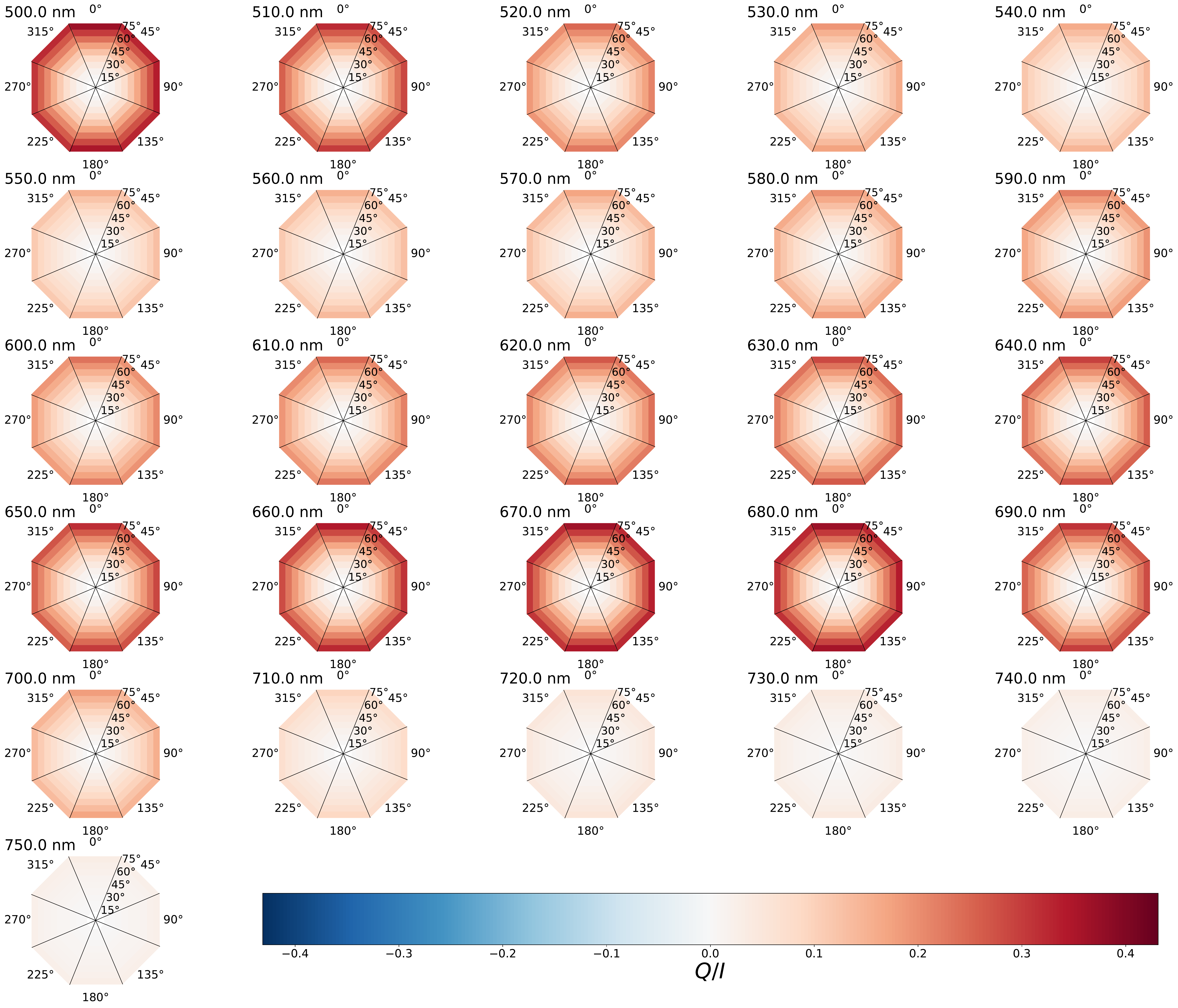}
	\caption{Polar plots per wavelength of the normalized Stokes parameter $Q/I$ for the average of 27 leaves. 
	The phase angle is given by the radial coordinate and the azimuth by the angular coordinate.}
	\label{fig:Qpol}
\end{figure}

\begin{figure}[!hb]
	\centering 
	\hspace*{-1.5cm}  
	\includegraphics[width=1.2\textwidth]{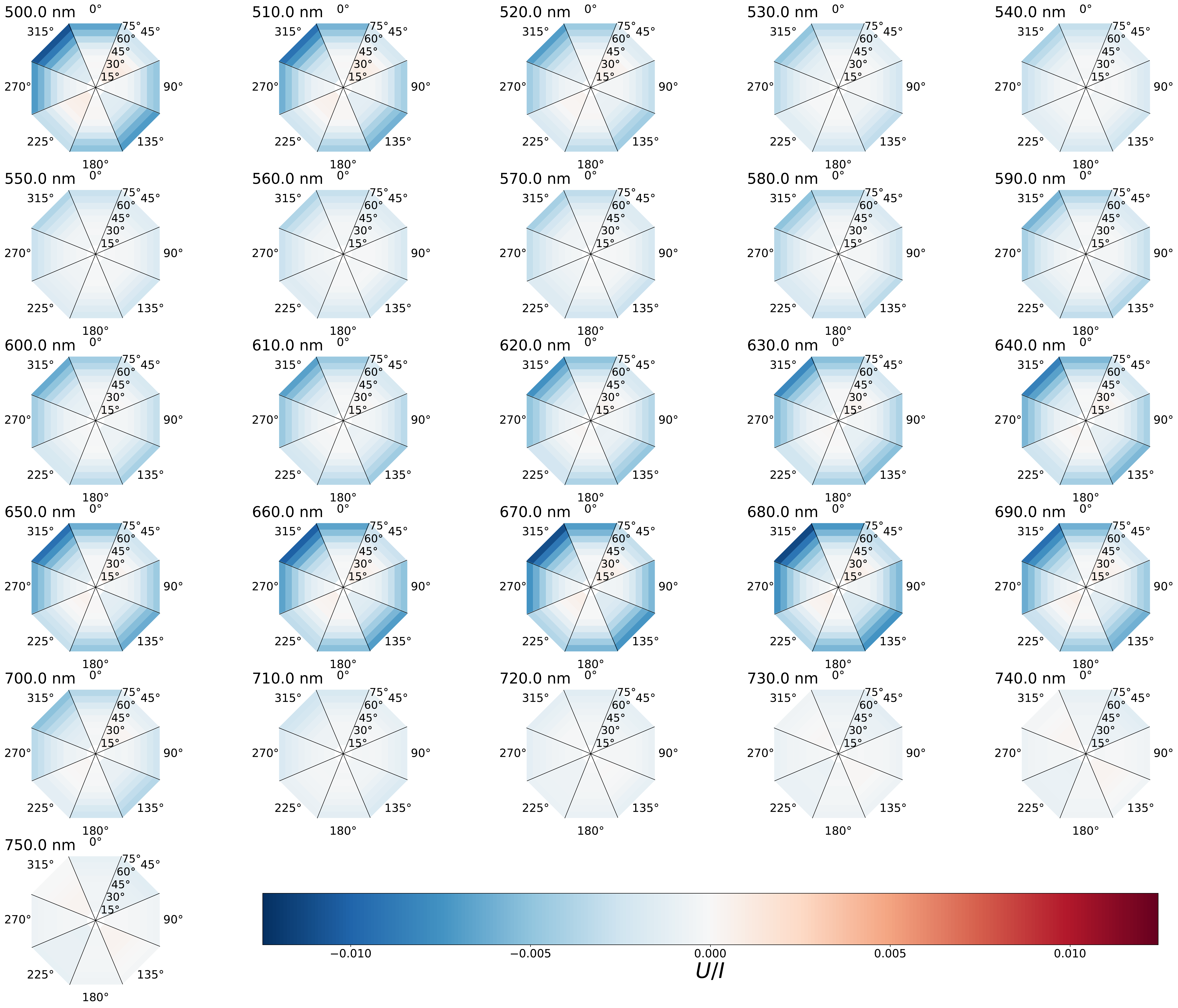}
	\caption{Polar plots of the normalized Stokes parameter $U/I$ per wavelength for the average of 27 leaves. The phase angle is given by the radial coordinate and the azimuth by the angular coordinate.}
	\label{fig:Upol}
\end{figure}

\begin{figure}[!hb]
	\centering 
	\hspace*{-1.5cm}   
	\includegraphics[width=1.2\textwidth]{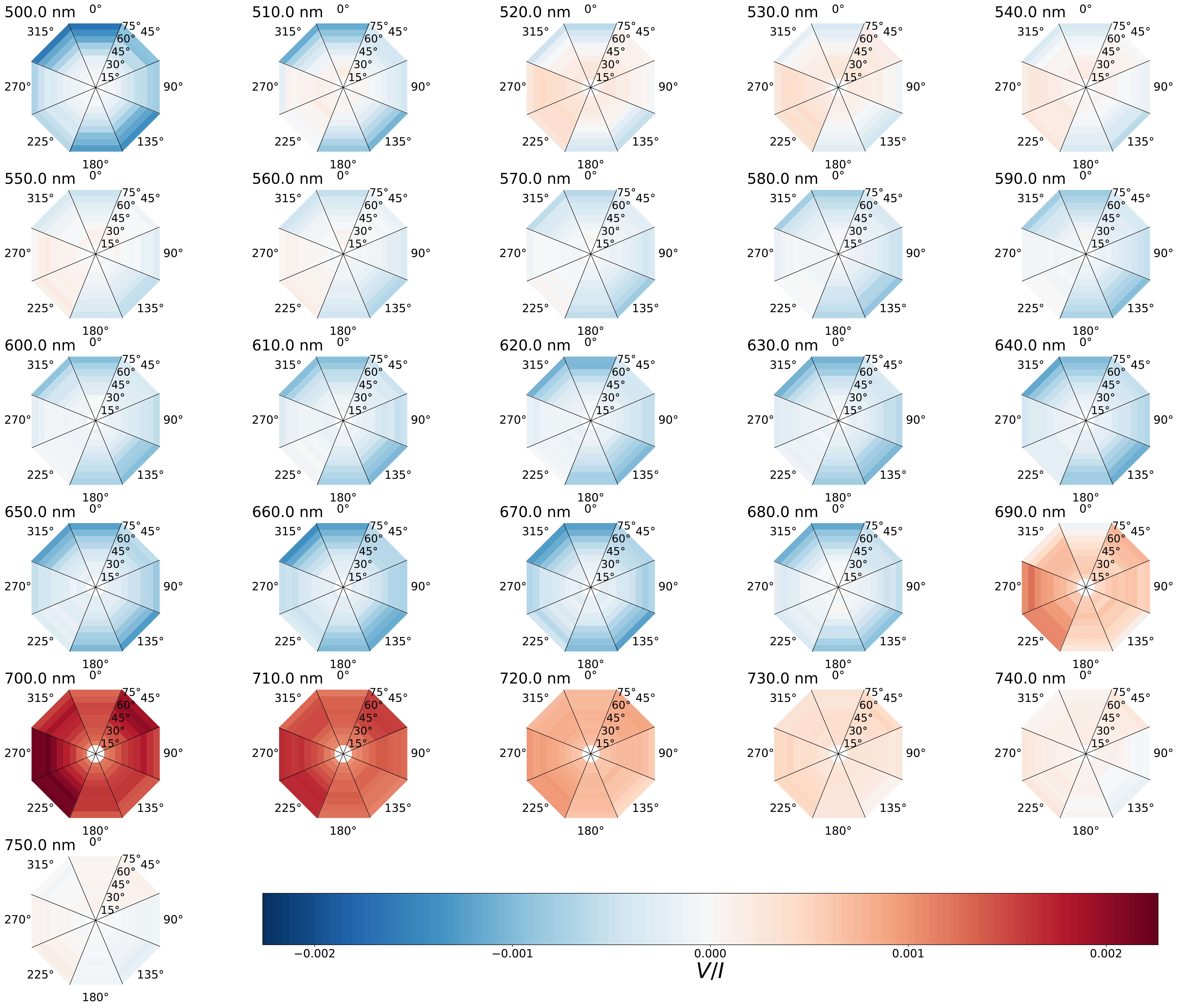}
	\caption{Polar plots of the normalized Stokes parameter $V/I$ per wavelength for the average of 27 leaves. The phase angle is given by the radial coordinate and the azimuth by the angular coordinate.}
	\label{fig:Vpol}
\end{figure}

All leaves were measured over a full $360^\circ$ rotation in steps of $45^\circ$. In Figure \ref{fig:Qpol} we show the non-azimuthal averaged measurements per wavelength and phase angle for Stokes $Q/I$. Figures \ref{fig:Upol} and \ref{fig:Vpol} show these results for $U/I$ and $V/I$ respectively. The 26 polar plots present the different wavelengths that were measured. For each plot, the radius $r$ represents the phase angle, while the azimuthal rotation is provided by $\varphi$. 

For $Q/I$ the variation in polarizance per azimuthal rotation is minimal although slightly increased with increasing phase angles. The difference in polarizance was the highest around the chlorophyll absorbance band (670 nm). For a phase angle of $75^\circ$, the polarizance at an azimuth of $0^\circ$ and $180^\circ$ was $0.43$, while this was $0.37$ for an azimuth of $45^\circ$ and $225^\circ$. At smaller phase angles no significant differences were observed.

For $U/I$ the differences in polarizance per azimuthal rotation were larger compared to $Q/I$ as is shown in Figure \ref{fig:Upol}. On average, the polarizance was the highest at an azimuth of $135^\circ$ and $315^\circ$, while smallest at an azimuth of $45^\circ$ and $225^\circ$. For example, at a phase angle of $75^\circ, U/I$ was -0.012 at $315^\circ$ and only $-1.9*10^{-3}$ at $45^\circ$. At an azimuth of $45^\circ$ and $225^\circ$, Stokes $U/I$ was positive at lower phase angles ($10-37.5^\circ$). For low phase angles ($10-30^\circ$) this results in an azimuthal average with no significant polarization signals (see also Figure \ref{fig:Stokesgraphim}). For many individual species, also at higher phase angles both positive and negative $U/I$ was observed depending on the azimuthal rotation. This results in a per species $U/I$ signal, averaged over azimuth, that  also at high phase angles is not always significant (see also the appendix for the rotational averages of the individual species).

Stokes $V/I$ (see also Figure \ref{fig:Vpol}) shows small variations in polarizance per azimuthal rotation. While at lower phase angles there are no significant variations, at phase angles above $45^\circ$ the polarizance was more slightly more negative at an azimuth of $0^\circ$ and $180^\circ$ and at an azimuth of $135^\circ$ and $315^\circ$. On the other hand, the polarizance was more slightly more positive at an azimuth of $45^\circ$ and $225^\circ$ and at an azimuth of $90^\circ$ and $270^\circ$. For $V/I$ these variations seem primarily related to shifts in the zero line. When the difference between the positive and negative psi-type band per azimuth is taken no significant variations between the azimuthal rotations are observed.

\begin{figure}[!htb]
	\centering 
	\hspace*{-2.5cm}  
	\includegraphics[width=1.5\textwidth]{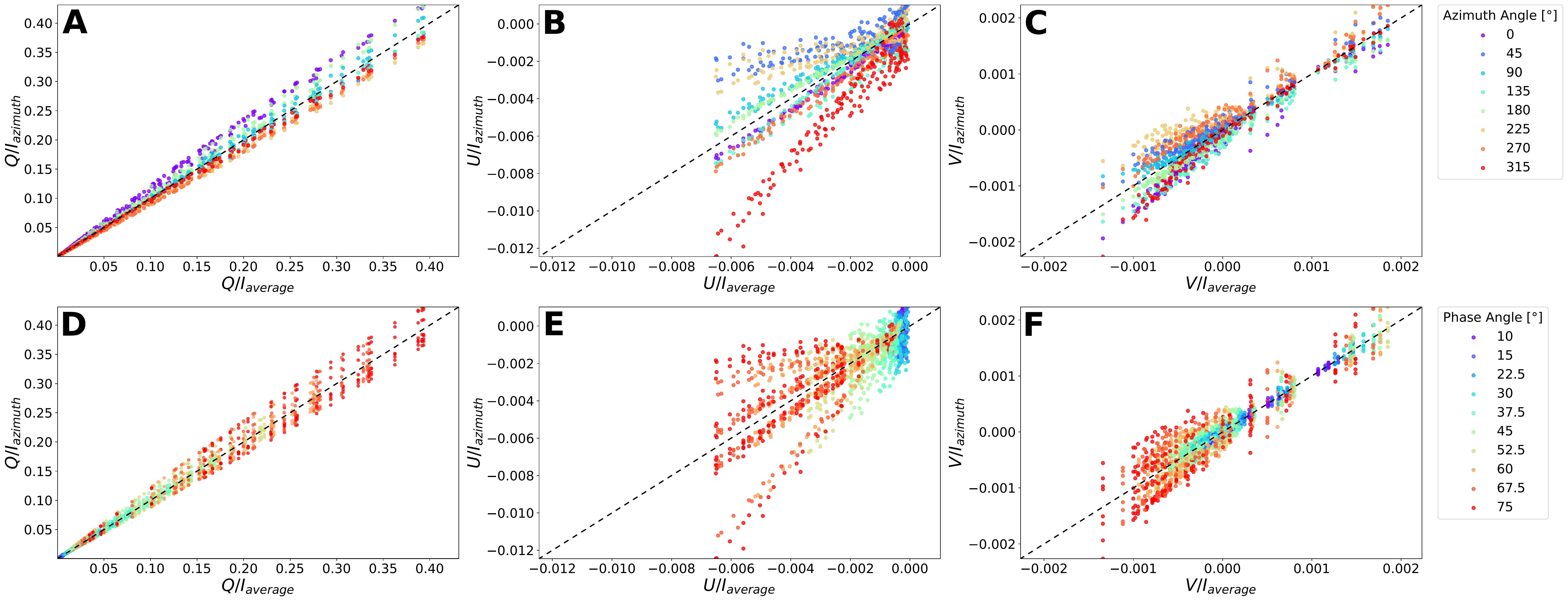}
	\caption{The normalized Stokes parameters $Q/I$ (\textbf{A} and \textbf{D}), $U/I$ (\textbf{B} and \textbf{E}) and $V/I$ (\textbf{C} and \textbf{F}) per leaf azimuthal measurement vs the leaf average over all azimuthal angles. In panels \textbf{A}, \textbf{B} and \textbf{C} the colors indicate the azimuth angle. Panels \textbf{D}, \textbf{E} and \textbf{F} show the same data but the colors indicate the different phase angles.}
	\label{fig:Scatteraz}
\end{figure}

In Figure \ref{fig:Scatteraz} we show the values for Stokes $Q/I$ (\textbf{A} and \textbf{D}), $U/I$ (\textbf{B} and \textbf{E}) and $V/I$ (\textbf{C} and \textbf{F}) per azimuth versus the average over all rotations. In the upper panels (\textbf{A},\textbf{B} and \textbf{C}) the color of the data points represent the azimuth angles, while in the lower panels (\textbf{D}, \textbf{E} and \textbf{F}) the same data is shown with the colors representing the phase angles. 

It is clearly visible that some azimuthal directions have a higher or lower polarizance than the rotational average. An extreme for a direction in one of the Stokes parameters does not lead to a similarly high or low value in the other parameters. The differences between the azimuthal directions and the average increase with increasing phase angles (Figure \ref{fig:Scatteraz} \textbf{D}, \textbf{E} and \textbf{F}) as is also shown in Figure \ref{fig:Stokesgraphim}. 

\subsection{Polarization Edge}

\begin{figure}[!htb]
	\centering 
	\hspace*{-0 cm}  
	\includegraphics[width=1.0\textwidth]{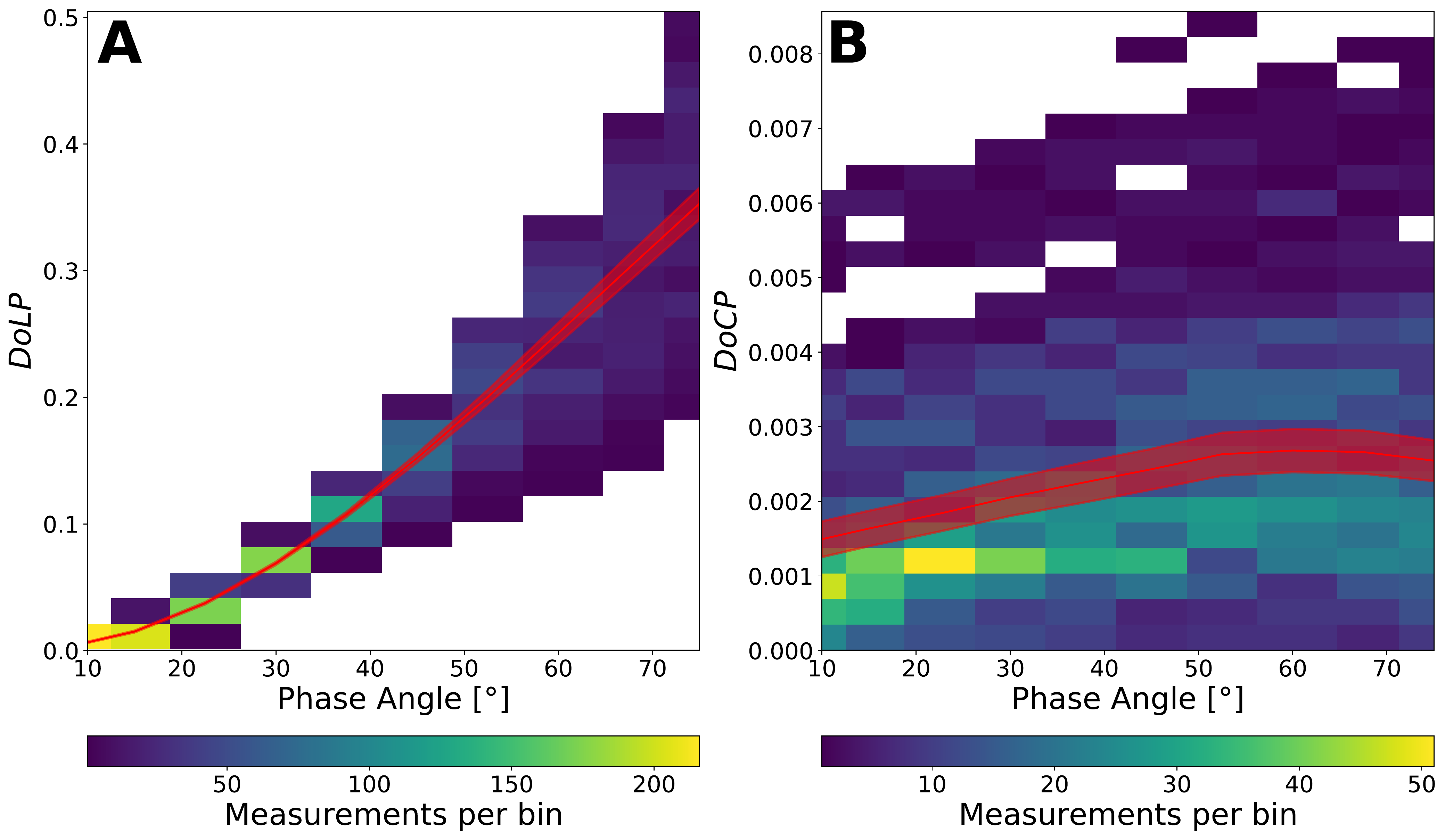}
	\caption{A 2D histogram showing the polarization edge of the linear polarization (DoLP) (\textbf{A}) and the circular polarization (DoCP) (\textbf{B}) versus the phase angle for all measurements. The red line represents the average of 27 different leaves and the red shaded area represents the standard error.}
	\label{fig:Poledge}
\end{figure}

Similar to the vegetation red-edge, which is the steep increase in scalar reflectance between the chlorophyll absorption band in the red (around approximately 670 nm, see Figure \ref{fig:Reflect}) and the region of maximum reflectance in the near-infrared (reached after approximately 750 nm, see Figure \ref{fig:Reflect}), the strength of the polarization signals can be described by their 'polarization edge'. We have grouped $Q/I$ and $U/I$ together in the degrees of linear polarization ($DoLP$) (i.e. $DoLP = \sqrt{Q^2+U^2}/I$) which makes more sense from a biosignature point of view. For the $DoLP$ we have measured the polarization edge by taking the value at 670 nm, where the polarizance is maximal for both $Q/I$ and $U/I$, and subtracting it with the background polarization at 750 nm. For the circular polarization, we compute the polarization edge as the difference between the negative (670 nm) and positive (700 nm) psi-type circular polarizance band, where the absolute value is expressed as the degrees of circular polarization ($DoCP$). Note that we did not correct for the background polarization at 750 nm as due to the split signal this will only lead to a total offset in $V/I$ but not a change in magnitude. Additionally, the magnitude at 750 nm was less than $5*10^{-5}$ (see also Figure \ref{fig:Stokesmaxmin} (\textbf{C}). We show the polarization edges in Figure \ref{fig:Poledge} of the linear polarization \textbf{A} and of the circular polarization \textbf{B} shown in degrees of circular polarization. The red line represents the average of 27 leaves while the edges per measurement are grouped into bins, with the amount per bin represented by the colorbar.

\begin{figure}[!thb]
	\centering 
	\hspace*{-0 cm}  
	\includegraphics[width=1.0\textwidth]{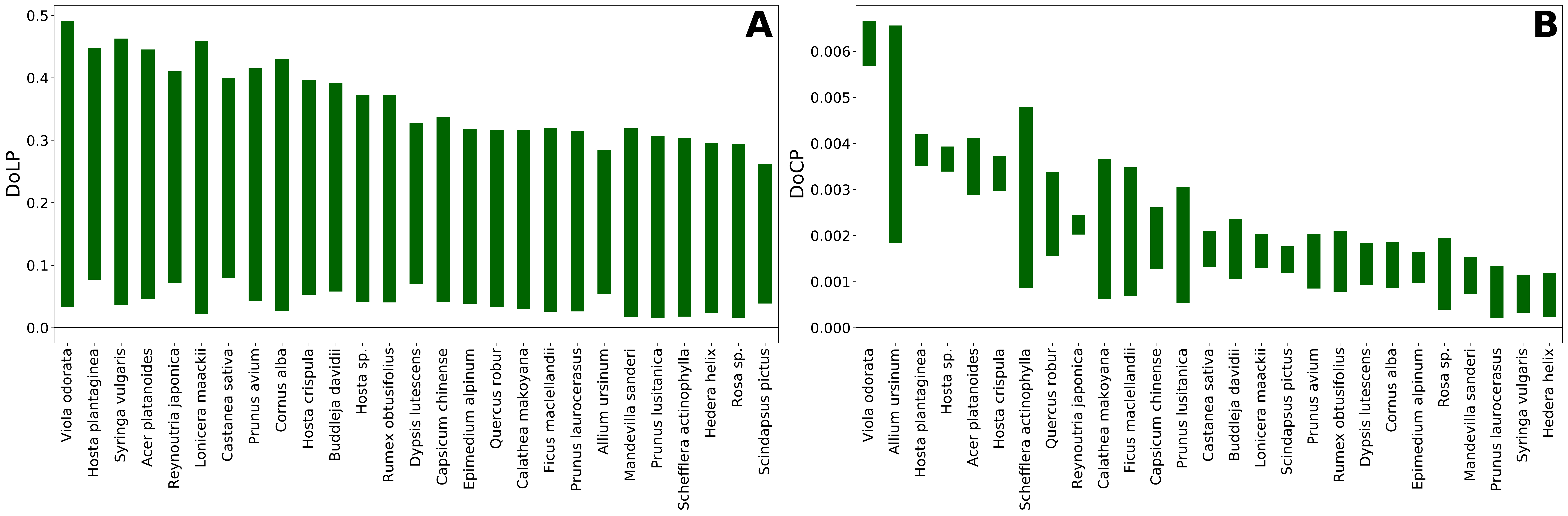}
	\caption{The maximum and minimum polarization edge over phase angle of the linear polarization (DoLP) (\textbf{A}) and the circular polarization (DoCP) (\textbf{B}) for all measured species averaged over azimuth. Data are sorted on the average edge value.}
	\label{fig:PoledgeSp}
\end{figure}

For the $DoLP$ the largest difference between the minimum and maximum polarizance was observed at $75^\circ$, where the edge had a magnitude of $0.35$. As the maximum polarizance was still increasing with increasing phase angles it is likely that the peak polarizance and edge magnitude occurs at even higher phase angles. For the $DoCP$, the maximum edge was measured at $60^\circ$, with an average value of $2.7*10^{-3}$. We show the maximum and the minimum extend of the polarization edges per species in Figure \ref{fig:PoledgeSp}. 

\subsection{Non-vegetation polarizance}

\begin{figure}[!thb]
	\centering 
	\hspace*{-1.5cm}  
	\includegraphics[width=1.2\textwidth]{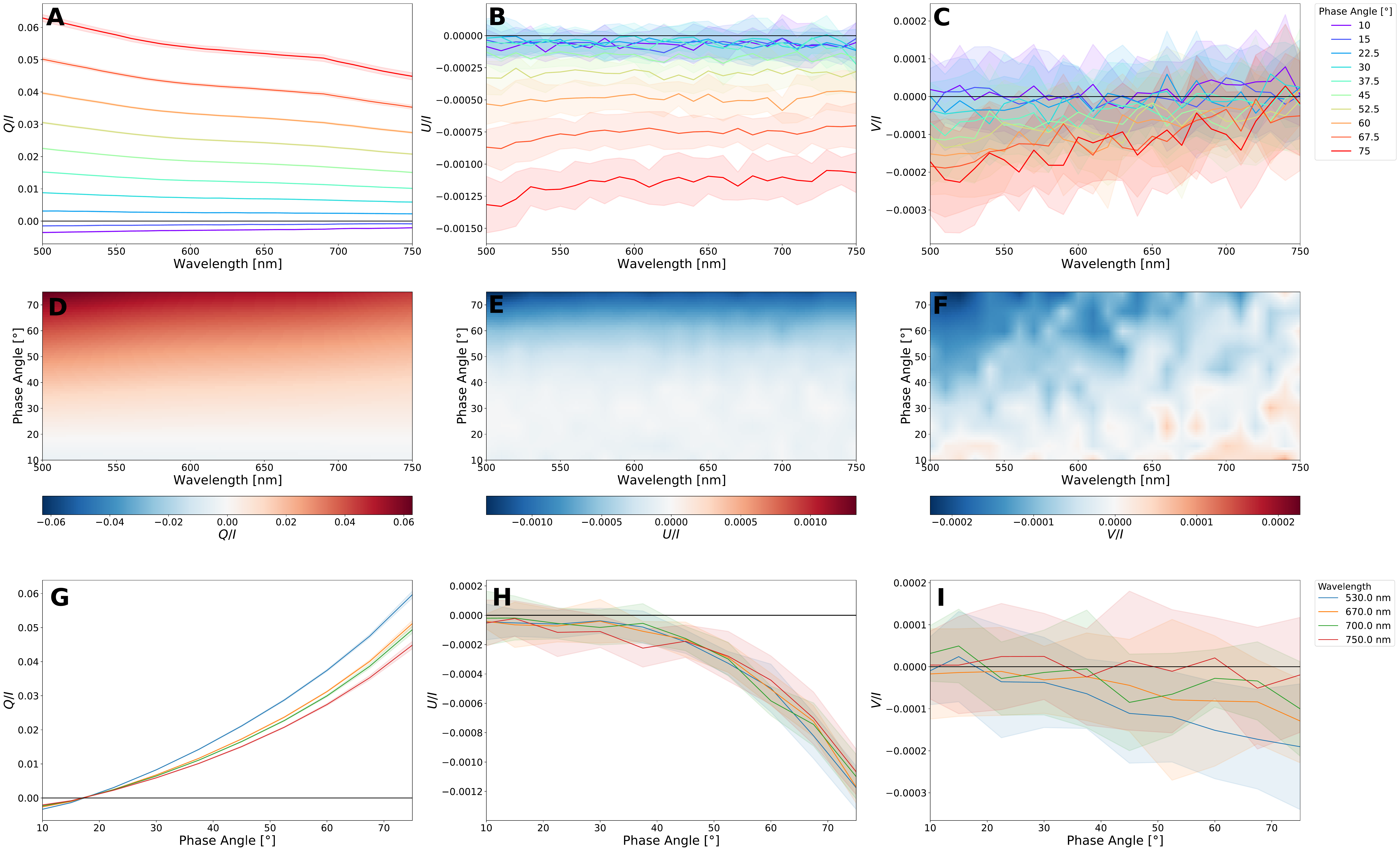}
	\caption{The average polarization spectra $Q/I$ (\textbf{A},\& \textbf{D}) $U/I$ (\textbf{B} \& \textbf{E}) and $V/I$ (\textbf{C} \& \textbf{F}) of 3 samples of sand averaged over azimuth. In the upper panels, the colors represent the different phase angles and the shaded areas denote the standard error. In the bottom panels, the colors represent different wavelengths and the shaded areas denote the standard error.}
	\label{fig:Sand}
	
\end{figure}

For comparison with the vegetation polarization data, we show in Figure \ref{fig:Sand} the linear (\textbf{A} and \textbf{B}) and the circular polarization spectra (\textbf{C}) versus the phase angle of 3 sand samples. Again, Figure \ref{fig:Sand} (\textbf{C},\textbf{D} and \textbf{E}) represents the same data visualized in a 2-D display. Selected wavelengths, similar to those in Figure \ref{fig:Stokesmaxmin}, versus the phase angle are shown in \textbf{G}, \textbf{H} and \textbf{I} for $Q/I$, $U/I$ and $V/I$ respectively.

Similar to vegetation, $Q/I$ strongly increases with increasing phase angles. Also for sand, the linear polarizance correlates to its reflectance spectrum which in the visible gradually increases with increasing wavelength. At small phase angles (see also Figure \ref{fig:Sand} (\textbf{G})), $Q/I$ was negative, thus the polarization was oriented parallel to the plane of scattering. $Q/I$ changed sign at $\sim17^\circ$ for all measured wavelengths. We observed a similar negative $Q/I$ at small phase angles (typically below $22.5^\circ$) for some leaves, but on average $Q/I$ was positive for all measured wavelengths (see Figure \ref{fig:Stokesgraphim} (\textbf{A})).

Stokes $U/I$ is approximately 0 for all wavelengths below a phase angle of $37.5^\circ$ but then increases in overall with increasing phase angles and is spectrally featureless. The circular polarizance, too, is not showing any spectral features. Especially towards the blue $V/I$ increases slightly with increasing phase angles, with a maximum of $-2.2*10^{-4}$.

\section{Discussion and conclusions}
We present in this study the systematic investigation of full-Stokes spectropolarimetry on leaves and its dependency on the phase angle. Importantly, while some studies and models exist for the linear polarizance of vegetation, studies showing the relationship between vegetation circular polarizance and phase angle are lacking. It has been demonstrated that circular polarimetry is able to provide an unambiguous means to detect homochiral matter, considered an important universal agnostic biosignature \cite{Sparks2009,Patty2018a}. The results in this study demonstrate that the characteristic spectropolarimetric properties of vegetation remain relatively unchanged with changes in measurement geometry, underlining the potential of circular spectropolarimetry as a means of detecting life beyond Earth.

Accounting only for the molecular phenomena, the vegetation circular polarizance should in principle be insensitive to phase angle variations. While the phase angle is important at the molecular scale and in oriented isolated constituents, which can result in very different spectral shapes \cite{Garab1991,Finzi1989}, it can be expected that these variations even out at larger scales where the chloroplasts are more randomly distributed. Leaves, however, are geometrically complex structures with many different cellular interfaces and covered by a extracellular lipid structure with varying chemical compositions that can all potentially affect its spectropolarimetric response \cite{Vanderbilt1991, Grant1987}. 

We demonstrate in the present study that the circular polarizance of vegetation is relatively insensitive to alterations in the phase angle. These results confirm previous circular spectropolarimetric field observations of vegetation canopy under overcast and cloudless sky conditions, showing a circular polarization spectrum that was virtually identical \cite{Patty2019}. Still, the magnitude of the signal varied with changes in phase angle, where the largest differences between the negative and positive band were observed at a phase angle of $60^\circ$ and the smallest at a phase angle of $10^\circ$ (Figure \ref{fig:Poledge}). The spectral shape of the typical split psi-type polarization, however, remained preserved. As such, integrating over a planetary disc should not yield spectral changes due to possible phase angle dependency. Additionally, it has been suggested that the vegetation circular polarizance could be a powerful tool in assessing vegetation physiology remotely \cite{Patty2017, Patty2021}. Remote sensing using circular spectropolarimetry will be able to directly probe the macromolecular structure of the photosynthetic apparatus and thus its functioning \cite{Patty2017, Lambrev2019} and as such could provide a powerful complementary tool in assessing effects of climate change and the monitoring of vegetation physiological state in vulnerable regions \cite{Patty2021}.

It should be noted that in some cases significant spectropolarimetric differences between species can be observed. For instance, in a \textit{Hosta} cultivar (see Appendix) we observed a negative psi-type polarization band with a magnitude of well over $-5*10^{-3}$ but with a postive band of only $1*10^{-3}$. On the other hand we measured for a \textit{Viola} cultivar (see Figure \ref{fig:Species}) a positive band with a magnitude of $6*10^{-3}$ but with a corresponding negative band less than $-1*10^{-4}$. Importantly, the two psi-type bands have a different physical origin, where the negative band is preferentially associated with the stacking of the thylakoid membranes inside the chloroplasts, and where the positive band is mainly associated with the lateral organization of the chiral macro-domains \citep{Lambrev2019, Garab2009}. Similar differences, where some species have much larger negative or positive bands, have been reported before, see e.g. \cite{Patty2018c, Patty2021}. While the circular polarizance does not always provide a specific species signature and it will thus be difficult to determine the species based on the signal alone, it could potentially aid in enhancing the contrast in existing distinguishing techniques.

The linear polarizance of vegetation, too, can serve as a biosignature \citep{Berdyugina2016, Klindzic2021}. Importantly, the linear polarization created by vegetation could be used to enhance the contrast between the reflected light of an exoplanet and the very bright emission by its host star \cite{Klindzic2021} As has been reported before (see e.g. \cite{Peltoniemi2015, Vanderbilt1991,Vanderbilt1985a}), vegetation linear polarization is strongly dependent on the phase angle. At a phase angle of $10^\circ$, we measured an average linear polarization of $6.4*10^{-3}$. At $75^\circ$ the linear polarizance was much higher with a value of $0.35$ which on average continued to increase. Likely the maximum average polarizance lies between $75^\circ$ and two times the Brewster angle (which for leaves is $55.4^\circ$ on average \cite{Jacquemoud2019}).

In contrast to vegetation circular polarizance, the linear polarizance is more dependent on the pigments present in the leaf, rather than their molecular organization. The structural geometry of leaves can vary greatly between species and is not homogeneous. Some leaves have a thick shiny cuticle while others are abundant in trichomes (fine epidermal outgrowths or hairs). Importantly, a leaf surface is generally far from optically smooth \cite{Grant1987}. Larger structures are shown along the veins which can undulate from the surface but many plants have surfaces with angular aspects in multiple directions. Such phenomena can affect the obtained leaf linear polarizance \cite{Grant1987,Vanderbilt1985a}. Additionally, the interior anatomy of leaves can vary greatly between species, and can vary between leaves of the same species. While the effects of these variations on the polarization response has not been studied in depth, it is clear that they can effect the optical properties greatly and can have a large effect on scalar reflectance and transmittance \citep{Jacquemoud2019}. It has additionally been demonstrated that linear polarization can be informative of plant physiological status \citep{Vanderbilt2017,Vanderbilt2019, Yao2020}.

Similar to the scalar reflectance, edges mimicking biotic matter can readily emerge in the spectropolarimetry of abiotic matter, whereas circular polarizance is less prone to such confounding factors. Consequently, while circular polarization signals of comparatively small magnitude (but can be up to a couple percent for certain eukaryotic algae \citep{Patty2018c}), it has the potential to deliver a more robust and specific biosignature.
The scalar reflectance, linear polarizance and circular polarizance are, however, very complementary tools. In the end, a confident detection of extraterrestrial life candidates will presumably have to consist of a vast range of spectroscopic techniques. 

Overall, we have successfully demonstrated that the circular polarization biosignatures of vegetation are relatively insensitive to changes in phase angle, providing a promising outlook on future life detection or remote sensing applications. It will be important to compare the results in this study with controlled measurements of vegetation taken in the field. Additionally, it will be also key to incorporate the circular polarizance-phase angle relationship in (exo)planetary models with realistic surface and atmospheric components \citep{Groot2020}. While we have only measured vegetation in this study, more primitive phototrophs also show remarkable polarization features \cite{Sparks2021}. Although it is not expected that measurements on these organisms yield different results for the circular polarization phase angle dependency, such study would be highly interesting and important in understanding the nature of especially circular polarization as a biosignature.  

\section*{Acknowledgments}
This work has been carried out within the framework of the National Centre of Competence in Research (NCCR), PlanetS, supported by the Swiss National Science Foundation (SNSF).
\nolinenumbers
\clearpage

\bibliography{Directional_LP.bbl}
\bibliographystyle{unsrt}

\clearpage

\section{Supplementary Material}
\includepdf[pages=-]{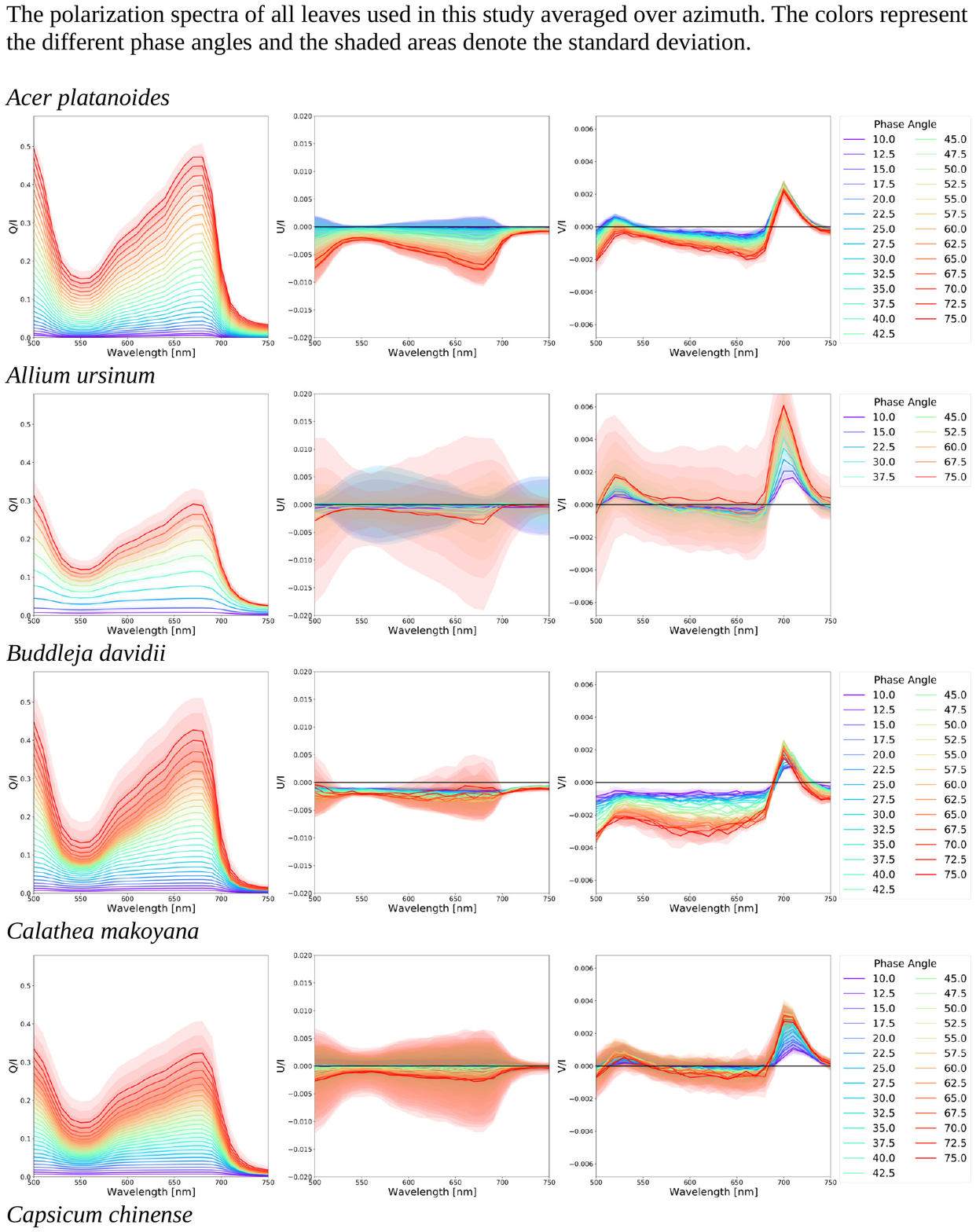}
\end{document}